\documentclass[prX,showpacs,superscriptaddress,twocolumn ]{revtex4}

\usepackage{graphicx}
\usepackage{amsmath}
\usepackage{amsfonts}
\usepackage{amssymb}
\usepackage{epsf}
\usepackage{hyperref}

\newcommand{\eps}[0]{\varepsilon}

\begin{document}
\title{Matter wave scattering on an amplitude-modulated optical lattice}
\author{P. Cheiney, C. M. Fabre, F. Vermersch, G. L. Gattobigio, R. Mathevet, T.~Lahaye and D.~Gu\'ery-Odelin}
\affiliation{Universit\'e de Toulouse, UPS, Laboratoire Collisions Agr\'egats R\'eactivit\'e, IRSAMC; F-31062 Toulouse, France}
\affiliation{CNRS, UMR 5589, F-31062 Toulouse, France}

\date{\today}

\begin{abstract}
We experimentally study the scattering of guided matter waves on an amplitude-modulated optical lattice. We observe different types of frequency-dependent dips in the asymptotic output density distribution. Their positions are compared quantitatively with numerical simulations. A semiclassical model that combines \emph{local} Floquet-Bloch bands analysis and Landau-Zener transitions provides a simple picture of the observed phenomena in terms of elementary \emph{Floquet photon} absorption-emission processes and envelope-induced reflections. Finally, we propose and demonstrate the use of this technique with a bichromatic modulation to design a tunable sub-recoil velocity filter. Such a filter can be transposed to all species since it does not rely on a specific internal level configuration of the atoms.
\end{abstract}


\maketitle

\section{Introduction}

Cold atoms interacting with time-modulated optical lattices display a wide variety of quantum and classical dynamics. These include the observations of dynamical localization \cite{KOS98,RSG00}, chaos-assisted tunneling \cite{HHB01,SOR01}, the Anderson metal-insulator transition in momentum space \cite{CLG08}, dynamically controlled tunneling \cite{LCS07,KSS08,CNA11}, matter wave engineering \cite{AFI10,ArH11,CrS11,SPP12} or the probing of gapped modes in degenerate quantum gases \cite{JSG08,HTA08}.

In this article we propose and demonstrate the use of time-dependent optical lattice for atom optics through the investigation of the scattering of a cold atom packet on an amplitude-modulated optical lattice \cite{Richter}. Our technique provides a new tunable tool for velocity selection in confined geometries and a system for studying quantum transport with time-dependent potentials \cite{BuL82} as initially studied in mesoscopic physics \cite{BWH98,PlA04}.

The Bragg reflection of a propagating guided matter wave on a static optical lattice has been recently demonstrated
\cite{FCG11}. The Bragg condition on the class of velocities that are reflected reads $v_{\rm{Bragg}}=nv_{\rm{L}}/2$ where $n$ is an integer, $v_{\rm{L}}=h/(md)$ and $d$ is the lattice spacing. This condition is valid in the perturbative limit i.e. for a small-depth optical lattice $U_0 \ll E_L$ where $E_L\equiv mv^2_L/2$ is the lattice energy scale. However $v_{\rm{Bragg}}$ is directly related to the periodicity of the lattice and therefore cannot be easily tuned over a large range.

In contrast, the interaction of a propagating matter wave with an amplitude-modulated optical lattice offers more flexibility and realizes, in particular, a tunable Bragg reflector. Our study explores the non perturbative regime for which the lattice depth is on the order of the lattice energy scale $E_L$  and the modulation depth is relatively large. In this range of parameters and for our typical interaction time, the stroboscopic phase space does not exhibit any chaos \cite{DLG} and the scattering can therefore be used to engineer the velocity distribution of the incoming packet of atoms.

The paper is organized as follows. In Secs.~\ref{experimentalsetup} and \ref{experimentalresults}, we present the experimental setup and review the main observations. In Sec.~\ref{vanishingdepthmodel}, we derive a simple model based on a modulated vanishing depth optical lattice. This approach is successfully compared with the experimental results. In Secs.~\ref{floquetblochframework} and \ref{semiclassicalmodel} we develop a semiclassical model that combines the local Floquet-Bloch framework and Landau-Zener transitions to get a deeper insight on the elementary processes at work in the matter wave scattering. In Sec.~\ref{velocityfilter}, we present an application with the realization of a tunable notch velocity filter through the scattering on an optical lattice whose amplitude is time-modulated by a bichromatic field. Conclusions are drawn in Sec.~\ref{conclusion}.

\section{Experimental setup}
\label{experimentalsetup}

\begin{figure}[!t]
   \begin{center}
      \includegraphics[width=7.5cm]{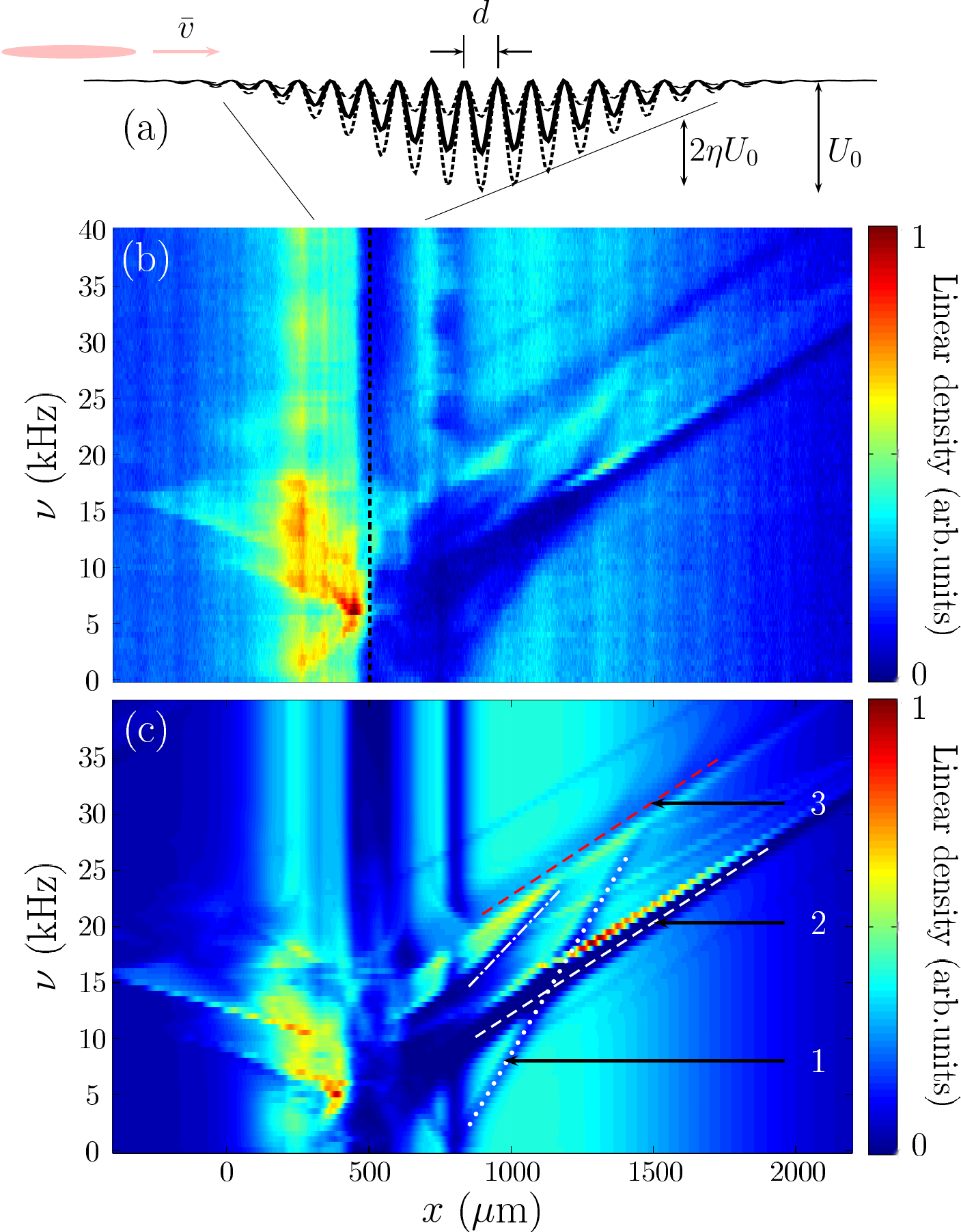}
      \end{center}
\caption{(color online) (a) Sketch of a propagating atomic packet impinging onto an optical lattice whose amplitude is modulated. (b) Measured longitudinal density $n(x,t_{\rm acc}+t_{\rm prop})$ after an acceleration stage of $t_{\rm acc}=15~\rm{ms}$ and a propagation time $t_{\rm prop}=78~\rm{ms}$ for various lattice modulation frequencies $\nu$ (lattice depth $U_0=2E_{L}$, modulation depth $\eta=33 \%$, lattice position given by the dotted line). (c) Numerical simulations (see text) with a resolution that matches the experimental optical resolution ($\sim10~\rm{\mu m}$).  Frequency-dependent dips are observed in the transmitted distribution.  Dotted, dot-dashed and dashed lines in (c) show the linear dependence of the dip position with $\nu$.}
\label{fig_exptheo}
\end{figure}

Our experimental setup has already been described in \cite{FCG11}. In short, a thermal cloud of typically $8.10^4$ rubidium-87 atoms at $T= 500$ nK is obtained after $3.5~\rm{s}$ of forced evaporation in a crossed dipole trap formed by two red-detuned (1070 nm) laser beams: a horizontal guide and a dimple beam. During the evaporation, we use the spin-distillation technique to prepare atoms in $\left| F=1, m_F= 0 \right\rangle$ \cite{CJK08,GCJ09}. We deliberately use a thermal cloud rather than a BEC in order to probe the modulated lattice for a wide range of velocities in a single shot (see below). By switching off the dimple beam,  we release a packet of longitudinal velocity dispersion $\Delta v=6~\rm{mm/s}$ in the horizontal guide. Atoms are subsequently accelerated  by a $t_{\rm acc}=15$ ms inhomogeneous magnetic field pulse to a mean velocity $\bar{v}=10~\rm{mm/s}$. The atomic packet then propagates towards the lattice whose center is located $500~\rm{\mu m}$ downstream from the trap position (see Fig.~\ref{fig_exptheo}(a)). The lattice is obtained by crossing two horizontal off-resonance laser beams (wavelength $\lambda=840~\rm{nm}$, waist  $w = 100~\rm{\mu m}$) at an angle $\theta = 81^{\circ}$ \cite{FCG11}. We modulate the lattice intensity using an acousto-optic modulator prior to the beam separation. The time-dependent potential experienced by the atoms reads:
\begin{equation}
U(x,t)=-U_0(t) e^{-2x^2/\tilde{w}^2}\sin^2\left( \frac{\pi x}{d}\right),
\label{potential}
\end{equation}
where  $U_0(t)=U_0\left(1+ \eta \cos\left(2 \pi \nu t\right)\right)$, $d= \lambda/ [2 \sin (\theta/2)]=650\pm 15~\rm{nm}$ ($v_{\rm{L}}=7.1$ mm/s, $E_L/h=\nu_L=5.4$ kHz) and $\tilde{w}=w/\cos(\theta/2)\simeq 130~\rm{\mu m}$. The lattice depth $U_0=2E_{L}$ is calibrated by Kapitza-Dirac diffraction \cite{OMD99}. The typical modulation depth is $\eta=33 \%$. The atomic packet propagates during $t_{\rm{prop}} =78~\rm{ms}$ through the lattice and is imaged \emph{in-situ} (without time-of-flight) by absorption imaging.

\section{Experimental results}
\label{experimentalresults}

Figure~\ref{fig_exptheo}(b) shows the atomic density after propagation as a function of the modulation frequency $\nu$. Each horizontal line is obtained by averaging 8 images integrated along the transverse direction. Two kinds of density dips are observed in the transmitted part: (i) dips whose positions do not depend on the modulation frequency and that correspond to velocity classes fulfilling the Bragg reflection on the \emph{static} lattice \cite{FCG11}, and (ii) dips whose positions depend on the frequency. As we shall discuss below, some dips of the latter category have their counterpart in the reflected packet and correspond to reflected class of velocity while others are due to slowing down or acceleration effects.

Except for the zones very close to depletion lines in the transmitted part in Fig.~\ref{fig_exptheo}(b), each position downward the lattice can be mapped onto a well defined class of incident velocity $x\simeq v_{\rm{inc}}(t_{\rm acc} +t_{\rm{prop}})+K$ where $K$ is a constant \cite{explicit}. Let us characterize the different depletion lines. Using the correspondence between $x$ and $v_{\rm inc}$, the main depletion line (white dashed line) in Fig.~\ref{fig_exptheo}(c) has a slope $(660~\rm{nm})^{-1}\simeq$ $1/d$. We also observe directly the corresponding reflected atoms in the region $x<0$. The red upper dashed line of depleted atoms in Fig.~\ref{fig_exptheo}(c) is parallel to the main line and has no counterpart in the reflected region. The white dot-dashed (dotted) line in Fig.~\ref{fig_exptheo}(c) has a slope twice (three times) as large as the one of the dashed white and red depletion lines in Fig.~\ref{fig_exptheo}(c).

 Figure \ref{fig_exptheo}(c) is the result of a numerical simulation of the atomic packet dynamics using the one-dimensional Schr\"odinger equation solved by the split-Fourier method and a wave packet whose initial momentum and position dispersions match the measured experimental values \cite{comment1}. We find a good agreement between simulations and experiment. However such an approach does not reveal the mechanisms responsible for the depletion lines. 
 
\section{The modulated vanishing depth model}
\label{vanishingdepthmodel}

\begin{figure}[!t]
   \begin{center}
      \includegraphics[width=8.cm]{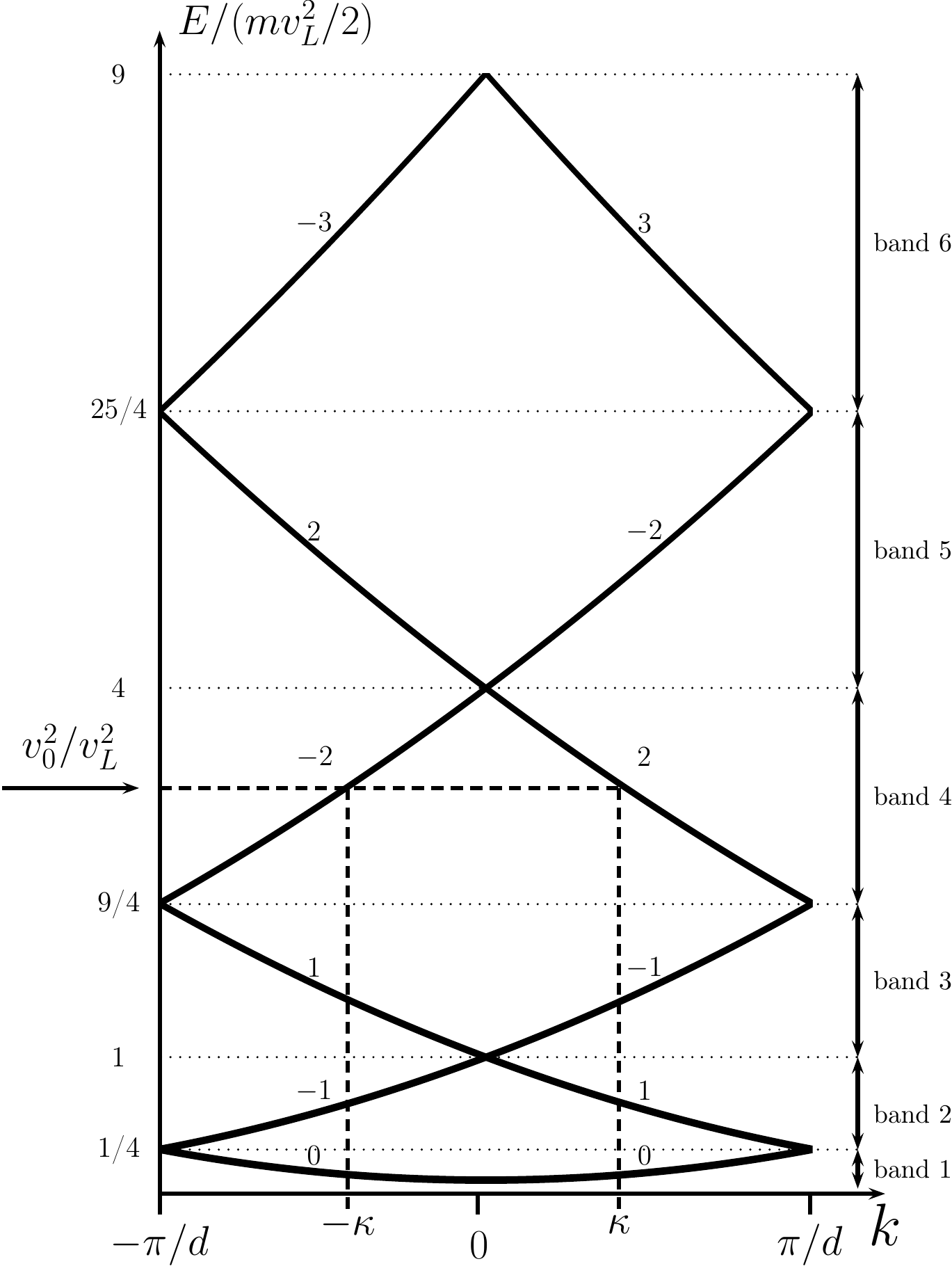}
      \end{center}
\caption{The band diagram is constructed by the superposition of parabolic energy spectra centered around all reciprocal points $E_n(k)=\hbar ^2(k-nk_L)^2/2m$. For each band we have indicated the index $n$ of the parabola from which the corresponding branches have been constructed. }
\label{fparab}
\end{figure}

The slopes and the relative position of the depletion lines can be simply interpreted in terms of interband transitions in the limit of a small lattice depth. Consider an incident quasi-monochromatic wave packet of velocity $v_0$. This velocity dictates the band in which the incoming atom enters. For instance, if $3v_{\rm{L}}/2 < v_0 <2v_{\rm{L}}$ the atom will be on band 4 (see Fig.~\ref{fparab}). 

In the limit of a vanishing lattice depth, the band structure can be constructed by the superposition of parabolic energy spectra centered around all reciprocal points $E_n(k)=\hbar^2(k-n k_L)^2/2m$, where $n$ is an integer and $k_L=2\pi/d$. For instance, bands 2 and 3 are constructed from the parabolas centered at $\pm \hbar k_L$.

\begin{figure}[h]
   \begin{center}
      \includegraphics[width=8.cm]{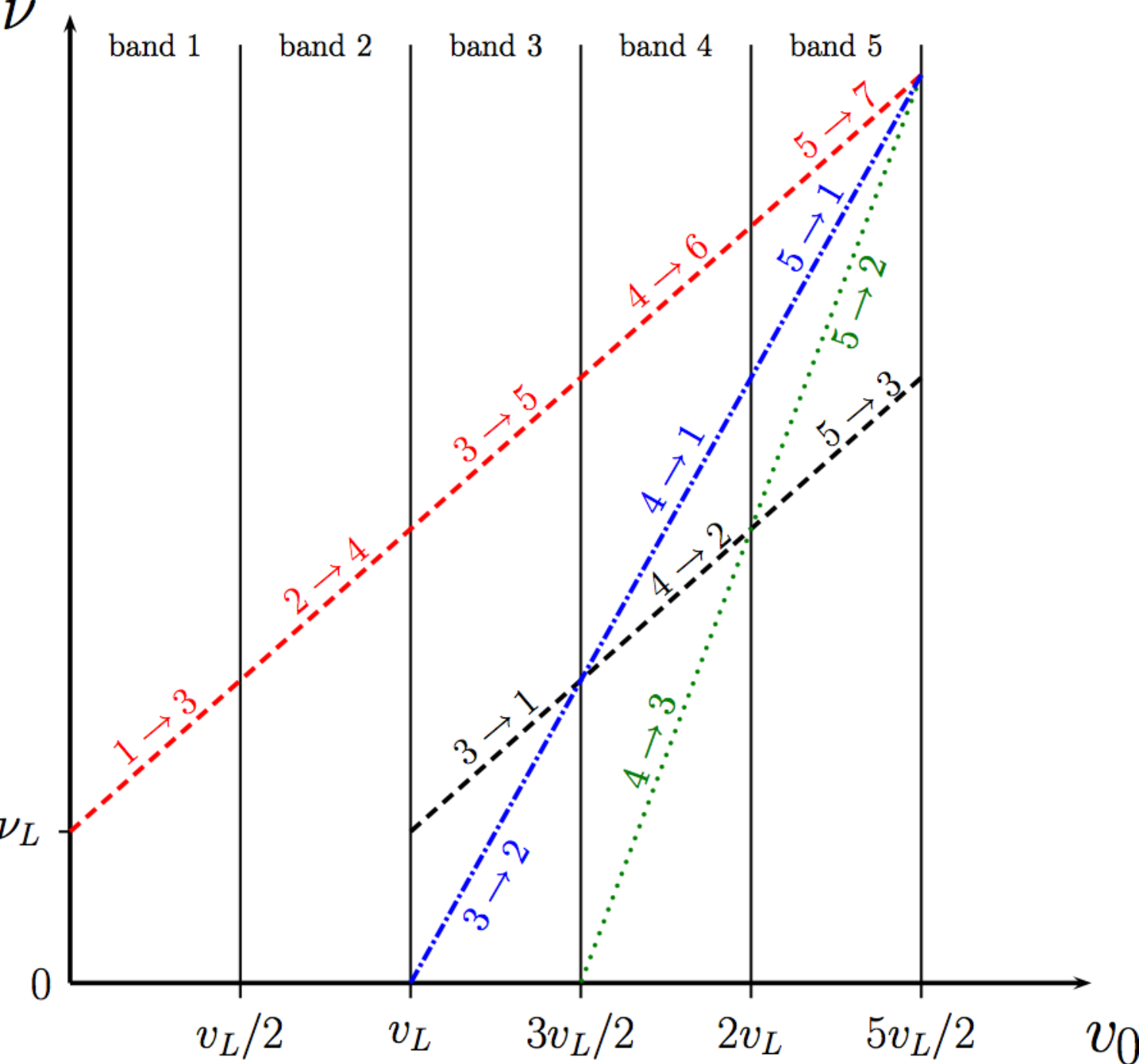}
      \end{center}
\caption{(color online) Schematics of the different depletion lines observed in Fig.~1 of the article along with the numbers of the bands involved in the transition dirven by the frequency of the amplitude modulation. With the experimental parameters, $v_{\rm{L}}\simeq 7.1$ mm/s.}
\label{fvit}
\end{figure}

In the limit of a small amplitude modulation, the modulation drives interband transitions that keep the pseudo-momentum $\hbar k$ unchanged.
In Fig.~\ref{fvit}, we have represented schematically the experimentally observed depletion lines in velocity space (see Fig.~\ref{fig_exptheo}). For each depletion lines and in the different velocity domains, we have also indicated two numbers: the first corresponds to the band on which atoms lies according to their incident energy $E=mv_0^2/2$ while the second corresponds to the band on which atoms are resonantly transferred by the modulation.

The interband transition frequencies that promote an atom from parabola $n$ to $n^\prime$ are given by
\begin{equation}
\pm \nu_{n \to n^{\prime}}=(E_{n}-E_{n^\prime})/h.
\label{smeq1}
\end{equation}
The sign $+$ $(-)$ corresponds to a transition to a lower (upper) band.
By energy conservation, we have
\begin{equation}
\frac{1}{2}mv_0^2 = E_{n}(\kappa).
\label{smeq2}
\end{equation}
Combining Eqs.~(\ref{smeq1}) and (\ref{smeq2}), we get
\begin{equation}
\pm \nu_{n \to n^{\prime}}=\frac{E_n-E_{n^{\prime}}}{h}=-(n-n^{\prime})^2\nu_L+\frac{n^{\prime}-n}{d}v_0.
\label{smeq3}
\end{equation}

As a first example, let us consider the green dotted depletion line (see Fig.~\ref{fvit}). In the incident velocity domain $3v_{\rm{L}}/2 < v_0 < 2v_{\rm{L}}$, it corresponds to a resonant transition between bands 4 and 3.
According to Fig.~\ref{fparab} we have \emph{a priori} two possibilities either a transition from parabola $n=2$ to parabola $n^\prime=-1$ or a transition from parabola $n=-2$ to parabola $n^\prime=1$ occurring at $\pm\kappa$. The slope of branch $n=-2$ is positive which thus corresponds to a propagation from left to right (positive group velocity). The states of the incoming packet with positive velocity in the range of energy that corresponds to band 4 will thus be projected onto the states that corresponds to the branch $n=-2$. Knowing $n$ and $n^\prime$ we deduce from Eq.~(\ref{smeq3}), the equation for the dotted line is $\nu=-9\nu_L+3v_0/d$.

Such an interpretation can be made for all depletion lines.
The lower (black) dashed line of Fig.~\ref{fvit} corresponds successively to the interband transitions $3 \to 1$, $4 \to 2$, $5 \to 3$ ...
From Eq.~(\ref{smeq3}), we deduce its equation $\nu=-\nu_L+v_0/d$.
Similarly we find for the upper (red) dashed line $\nu=\nu_L+v_0/d$. The offset between the two dashed lines is thus equal to $2\nu_L\simeq 10.8$ kHz. We obtain for the (blue) dot-dashed line the equation $\nu=-4\nu_L+2v_0/d$.


To get a better understanding of the width of the depletion lines, their interpretation in terms of elementary processes, the timescale on which the transitions occur and the role played by the Gaussian envelope of the lattice potential, we introduce now a more elaborated analysis based on Floquet-Bloch framework \cite{AsM76,Shi65}.

\section{The Floquet-Bloch framework}
\label{floquetblochframework}

\begin{figure}[!t]
   \begin{center}
\includegraphics[width=8.cm]{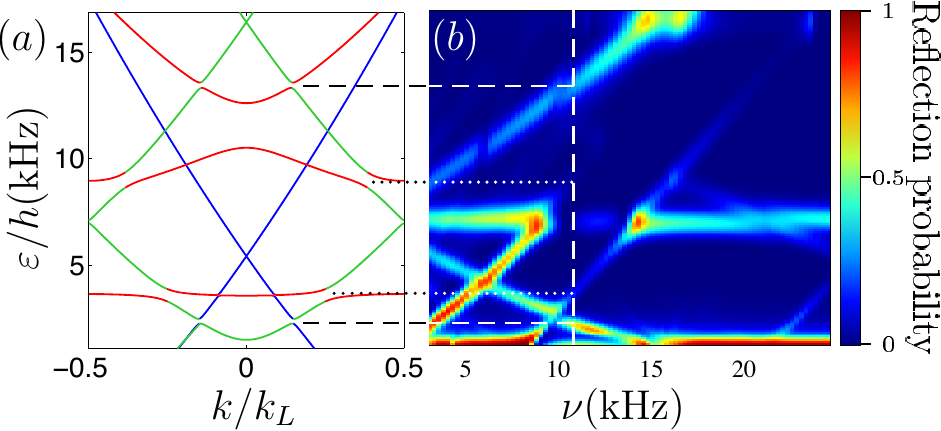}
\end{center}
\caption{(color online) (a) Floquet-Bloch band diagram for a square-envelope lattice (depth $U_0=2E_L$) and modulation frequency $\nu=11~\rm{kHz}$. Band color code: green  $-0.5 < \langle n_F\rangle < 0.5$, red $\langle n_F\rangle>0.5$ and blue $\langle n_F \rangle<-0.5$ where $n_F$ is the Floquet excitation number. (b) Probability of reflection obtained from a numerical simulation of a 1D wavepacket with  an incident velocity dispersion $\Delta v =0.2$ mm/s impinging on time amplitude-modulated lattice with a finite square-envelope (length=80 $d$, $U_0=2E_L$, $\eta=30\%$) as a function of the incident energy and $\nu$. The horizontal white dashed line shows the case $\nu=11~\rm{kHz}$ that corresponds to the diagram (a). The horizontal dashed (dotted) lines denotes open gap (degenerate) anticrossings. Only open gap anticrossings yield reflection. }
\label{simu_f_v}
\end{figure}

This approach is not restricted to small modulation depth and is thus well-adapted to analyze the experimental situation. For a potential periodic in both space and time, the Floquet-Bloch solutions of the time-dependent Schr\"odinger equation read:
\begin{equation}
\psi_{n,k}(x,t)=e^{i(kx-\eps_n(k)t/\hbar)} u_{n,k}(x,t),
\end{equation}
where $\eps_n(k)$ are the quasi-energies. The functions $u_{n,k}(x,t)$ are biperiodic in space and time and therefore can be Fourier expanded:
\begin{eqnarray}
u_{n,k}(x,t)=u_{n,k}(x+d,t)=u_{n,k}(x,t+T)\nonumber\\
=\sum_p \sum_{n_F} \phi^{n_F,p}_{n,k} e^{i(pk_Lx-{n_F}\omega t)}.
\end{eqnarray}
In the following, we restrict ourselves to $n_F \in \{-1, 0, 1\}$ i.e. to situations in which only one Floquet photon can be absorbed or emitted \cite{footnote2}.  At zero modulation depth, the Floquet-Bloch band diagram is nothing but the superposition of the Bloch diagrams shifted by $n_F\hbar \omega$. At finite modulation depth, anticrossings appear for frequencies that correspond to interband transitions.

Consider first the simple case of a square-envelope lattice of amplitude $U_0$ modulated at a frequency $\nu $ with an amplitude $\eta$. In Fig.~\ref{simu_f_v}(a) we plot the Floquet-Bloch spectrum for $\nu=11$ kHz. Two kinds of anticrossings can be identified: those yielding open gaps (horizontal dashed line in Fig.~\ref{simu_f_v})  and those without gaps for which two states with the same quasi-energy are available (horizontal dotted line in Fig.~\ref{simu_f_v}).  To identify the role of the different types of anticrossings on the incident matter wave packet, we have performed a 1D simulation that solves the corresponding time-dependent Schr\"odinger equation. Figure \ref{simu_f_v}(b) gives the reflection coefficient as a function of the incident energy $E_0$ and the modulation frequency $\nu$.
Two types of reflection conditions can be clearly identified: (i) those due to Bragg reflection onto the static lattice (no dependence on $\nu$) and (ii) those that correspond to open gap anticrossings and whose positions depend on $\nu$. The interpretation is clear: when the incident energy falls in an open gap anticrossing, no propagating state is available and the particle is reflected. The degenerate anticrossings do not induce reflection in the square-envelope case. However, as we discuss below, they play an important role in the dynamics of the experimentally relevant case in which the lattice has a slowly varying envelope.

\section{The semiclassical model}
\label{semiclassicalmodel}

In this latter case, the situation turns out to be different since the system can follow adiabatically a quasi-energy band during its time evolution. To describe this propagation, we propose a semiclassical model that enables one to identify the elementary processes responsibles for the velocity changes of the particle and the position at which such processes occur. It contains two main ingredients. The first one consists in describing the particle motion on a given \emph{local} Floquet-Bloch band through the combined evolution of the wavepacket position and of its mean pseudo momentum $k$. The corresponding set of  coupled equations reads
\begin{equation}
\dot{x} =\frac{1}{\hbar}\frac{\partial \eps_n}{\partial k} \qquad \mbox{and} \qquad \dot{k} =-\frac{1}{\hbar}\frac{\partial \eps_n}{\partial x}.
\label{eqmvt}
\end{equation}
The first equation defines the group velocity of the wave packets while the second results from the adiabatic following condition ${\rm d} \eps_n(k,x)/{\rm d} t=0$ \cite{AsM76,DrH99}. The second ingredient consists in taking into account the possibility for a particle to undergo a Landau-Zener transition when it passes an anticrossing. In our case, the approximation of a local two-level situation is valid and therefore the probability to change the band index is $P=e^{-2 \pi \gamma}$ with
\begin{equation}
\gamma= \frac{\Delta E^2}{4 \hbar} \left|\frac{{\rm d}}{{\rm d}t}(\eps_n-\eps_{n\pm1}) \right|^{-1}
\label{gammaeq}
\end{equation}
where $\Delta E$ is the size of the gap \cite{LaL77,Zen32}.

\begin{figure}[t!]
   \begin{center}
      \includegraphics[width=8.cm]{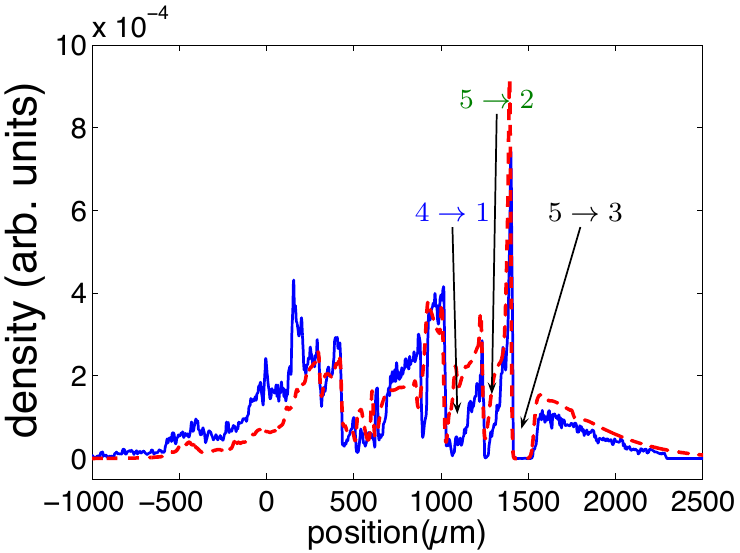}
      \end{center}
\caption{Dashed line: Result of the resolution of the 1D Sch\"odinger equation. Solid line, result of the semiclassical simulation with random Landau-Zener transition. The arrows indicate the main processes for three depletion lines. The wave packet has a mean velocity $\bar{v}=10~\rm{mm/s}$ and a velocity dispersion $\Delta v=6~\rm{mm/s}$. }
\label{comparaison}
\end{figure}

In practice, we evolve the particle according to Eqs.~(\ref{eqmvt}) and  we evaluate the energy difference to lower or upper bands at each time step. When this quantity reaches a minimum (i.e. at an avoided crossing position), we compute the
corresponding Landau-Zener probability and transfer or not the particle to the next band according to this probability.

To validate this semiclassical trajectory method, we compare it with the full resolution of the corresponding 1D Schr\"odinger equation. To perform this comparison, we have simulated the semiclassical trajectories of 1700 incoming velocities about the mean velocity of the packet in the following range $-4.5~{\rm mm/s}<v_{\rm{inc}}<24.5~{\rm mm/s}$.
In this way, we sample 98 \% of the initial distribution. Furthermore, we perform 35 shots for each incoming velocity to improve the statistics of our Monte Carlo simulation.

For each velocity class $v_{\rm inc}$, we get the density from the final positions of the different shots weighted by the initial wavepacket velocity distibution density at $v_{\rm inc}$. Figure \ref{comparaison} provides an example of such a comparison for $\nu=20~\rm{kHz}$.
The key features are very well captured by the semiclassical simulation. In this plot, we can easily identify the depletion lines that correspond to the different interband transitions.

\begin{figure*}[!t]
   \begin{center}
\includegraphics[width=13cm]{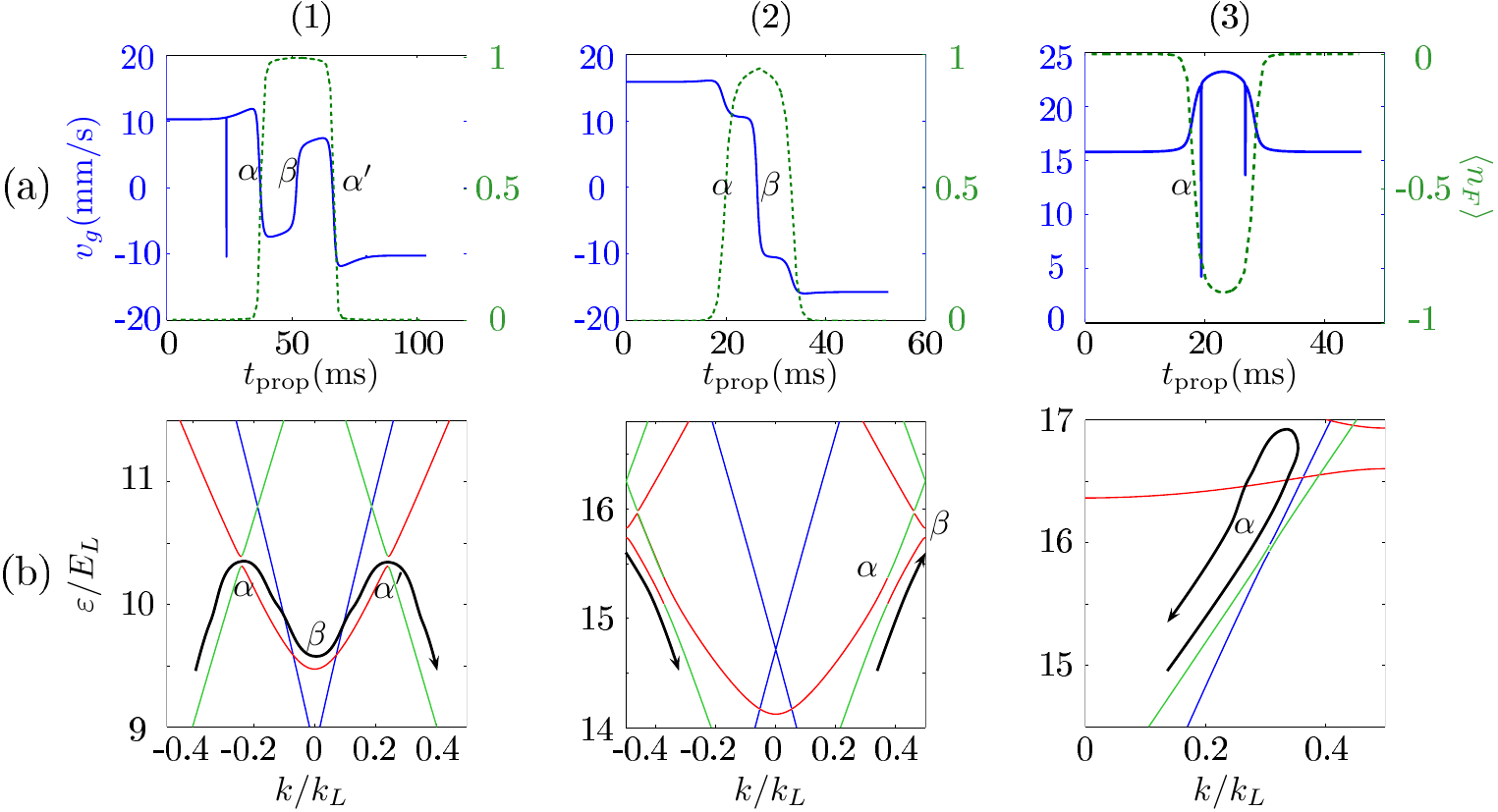}
\end{center}
\caption{(color online) (a) Velocity and mean Floquet excitation number $\langle n_F \rangle$ as a function of the propagating time for parameters corresponding to three different depletion zones shown in Fig.~\ref{fig_exptheo}. (b) Local Floquet-Bloch diagram. Dark arrow denotes the trajectory followed by the fictitious particle of the semiclassical model (see text). Case (1): Reflection on an open gap $v=10.3~\rm{mm/s}, \nu=11~\rm{kHz}$. Case (2): Reflection on an anticrossing without gap $v=15.8~\rm{mm/s}, \nu=20~\rm{kHz}$. Case (3) transient acceleration $v=15.8~\rm{mm/s}, \nu=30.5~\rm{kHz}$. $\alpha$ ($\beta$) denotes absorption or emission of one \emph{Floquet photon} (reflection).}
\label{stories}
\end{figure*}

We can use the semiclassical model to analyze the different mechanisms yielding to depletion bands as observed in Fig.~\ref{fig_exptheo}. For this purpose, we follow deterministically the branch for which the Landau-Zener transition probability is above 1/2 at an avoided crossing. To illustrate the wide variety of possibilities, we shall choose three generic and different set of parameters $(v_i,\nu)$ yielding to dips in the output density distribution (see labels 1, 2 and 3 in Fig.~\ref{fig_exptheo}(c)). In Fig.~\ref{stories}(a), we plot the velocity along with the mean Floquet excitation number for each case and for the main trajectory given by the Monte-Carlo simulation. In Fig.~\ref{stories}(b) we show the corresponding Floquet-Bloch diagrams in the region of interest. When a particle is moving toward the center, all quasi energies decrease since the amplitude of the attractive lattice increases. As a result, the particle state moves up relatively to the band diagram. In the same way, if the particle is moving backward, it will go down the hills of the diagram. With these simple pictures in mind trajectories can be readily interpreted. In each case, the key phenomenon is the absorption or stimulated emission of a Floquet photon by adiabatic following denoted by $\alpha$ in Fig.~\ref{stories}. In case (1), the particle emits a Floquet photon, performs a reflection (denoted $\beta$) when reaching the bottom of the band and absorbs a Floquet photon ($\alpha^\prime$) before leaving the lattice. In case (2), the first emission only slows down the particle which is then Bragg reflected and subsequently accelerated by Floquet photon absorption. In case (3), the particle is not reflected. It is transiently accelerated in the lattice by a Floquet photon absorption-emission cycle. For a much longer propagation time, the dip would be refilled.

Other features of the experimental and numerical diagrams of Fig.~\ref{fig_exptheo} can be readily explained thanks to our semiclassical model. For instance, the density bump above the white dashed line corresponds to atoms that have been slowed down.
The velocity spread of reflected particles at position 2 (see Fig.~\ref{fig_exptheo}) is 0.8 mm/s. This value can be recovered from the variation of the energy position of the gap along the lattice.

\begin{figure}[!t]
   \begin{center}
\includegraphics[width=9cm]{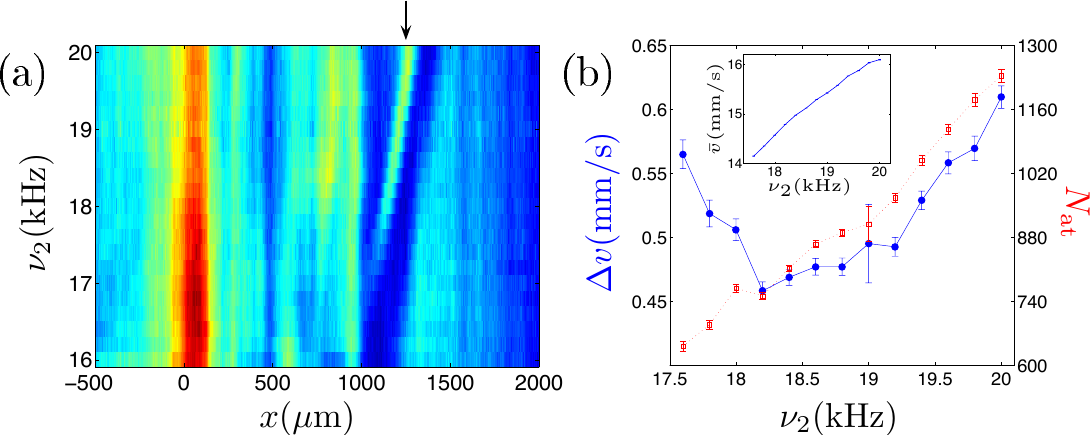}
\end{center}
\caption{(color online) (a) Density distribution for the scattering on a two-frequency modulated optical lattice (same conditions as in Fig.~\ref{fig_exptheo}) for a fixed frequency $\nu_1=16~\rm{kHz}$ and a scanned frequency $\nu_2$. A narrow slice of transmitted atoms is produced (see arrow). (b) Number of atoms and velocity dispersion associated with these narrow slices of transmitted atoms as a function of $\nu_2$. Inset: mean velocity of the slice of atoms as a function of $\nu_2$.}
\label{filtre}
\end{figure}

For a given incident kinetic energy $E_0$, a large size of the envelope and/or a large modulation depth increases the efficiency of the process since it favors an adiabatic following of the anticrossings.
A less intuitive feature concerns the lattice depth. Indeed, a small lattice depth ($U_0 < E_0$) increases the selectivity of the class of incident velocities that are affected by the modulation. This originates from the fact that the system is projected on a high energy band, and the position of the gap remains roughly constant throughout the lattice. Interestingly enough, this ensures the robustness of the method against the specific shape of a smooth envelope.

\section{Application: velocity filter}
\label{velocityfilter}

The narrowest velocity filters used in the cold atom community rely on velocity selective Raman transitions on atoms in free space. This technique involves a combined change of internal and external states. The achievable velocity width are in the range of 200-300 $\mu$m/s \cite{KaC92,SOR01,BCG04}. Using the scattering on amplitude modulated optical lattice we demonstrate hereafter a new technique to realize a velocity filter with a width slightly larger than the state of the art with velocity selective Raman transitions. Our technique uses only the external degrees of freedom and thus does not require any specific internal configuration. In addition it is well adapted for guided matter waves.

We turn our device into a tunable momentum filter by combining different modulation frequencies. We use here the main reflection line (white dashed line in Fig.~\ref{fig_exptheo}) that acts as a notch filter in momentum space. For this purpose, we modulate the lattice with two different frequencies to create a transmitted band between two rejected ones: $U_0(t)=U_0(1+\eta\cos(2\pi \nu_1 t)+\eta\cos(2\pi \nu_2 t))$.
Strictly speaking, the detailed dynamics of a wave packet submitted to this two frequency and non perturbative modulation cannot be inferred directly from the single frequency dynamics \cite{ChT04}. However, the simple picture according to which nearly independent dips can be drilled into the velocity distribution with two frequencies is quite robust. We observe that the reflection spectrum is roughly the product of the two independent spectra (see Fig.~\ref{filtre}) \cite{note}. The mean velocity of the slice of atoms is therefore governed by $d(\nu_1+\nu_2)/2$ while its width is controlled by the frequency difference $|\nu_2-\nu_1|$. In our set of experiments, $\nu_1$ is fixed at $16~\rm{kHz}$ and $\nu_2$ is varied from $16$ to $20$~kHz. Between the two reflection lines, atoms in a narrow class of velocity are transmitted (arrow in Fig.~\ref{filtre}). The slice contains about 1000 atoms and has a mean velocity on the order of 15 mm/s (inset of Fig.~\ref{filtre}). The minimum velocity dispersion of the velocity filter that we have designed is on the order of 450 $\mu$m/s for our parameters (i.e. 1.1 nK in temperature units).

\section{Conclusion}
\label{conclusion}

The matter wave engineering presented here does not have any fundamental limit. A further improvement of velocity selection could be achieved using a smaller depth lattice combined with a larger waist size.
This technique can be transposed easily to other species since it does not rely on a specific internal level configuration.
By construction, it is well adapted to 1D geometry and therefore enhances the toolbox of guided atom optics.
Finally, the control of the guided atomic flux for a given and tunable narrow class of velocities as studied here is reminiscent of the quantum modulated transistor principle where the gate voltage is replaced by the modulation \cite{SMR89}.

\begin{acknowledgments}

We thank useful discussion with R. Dubertrand, G. Lemari\'{e} and B. Georgeot.
We acknowledge financial support from the Agence Nationale pour la Recherche, the R\'egion Midi-Pyr\'en\'ees, the university Paul Sabatier (OMASYC project), the NEXT project ENCOQUAM and the Institut Universitaire de France.

\end{acknowledgments}


\begin{thebibliography}{99}
\bibitem{explicit} If $a$ denotes the acceleration experienced by the atoms during the magnetic pulse, one finds $K=at_{\rm acc}(t_{\rm acc}/2-t_{\rm prog})$.
\bibitem{comment1} In such a scattering experiment, atom-atom interactions are negligible in the course of the propagation.
\bibitem{footnote2} This assumption suffices to capture all observed phenomena on the experiment. The extension to larger Floquet excitations is straightforward.
\bibitem{explain}
The asymptotic values that enter Eq.~(\ref{gammaeq}) are obtained by a quadratic approximation of the energy difference $|\varepsilon_{n}-\varepsilon_{n\pm 1}|$. This ensures the robustness of the calculation of the Landau Zener probability.
\bibitem{note} The simple picture of two independent frequencies breaks down when the frequency difference is below 0.5 kHz.
\end{thebibliography}
\end{document}